\documentclass{iau}
\usepackage{graphicx}

\title[JD 11.~~ The size and mass evolution of massive galaxies] %% give here short title %%
{The size and mass evolution of the massive galaxies over cosmic time}

\author[Ignacio Trujillo]   %% give here short author list %%
{Ignacio Trujillo$^{1,2}$}

\affiliation{$^1$Instituto de Astrof\'{\i}sica de Canarias, c/ V\'{\i}a L\'actea s/n, 
E-38205, La Laguna, Tenerife, Spain  \\[\affilskip]
$^2$Departamento de Astrof\'{\i}sica, Universidad de La Laguna, E-38205, La 
Laguna, Tenerife, Spain\\ email: {trujillo@iac.es}}

\pubyear{2013}
\volume{295}  %% insert here IAU Symposium No.
\pagerange{xx--xx}
\jname{The intriguing life of massive galaxies}
\editors{D. Thomas, A. Pasquali \& I. Ferreras, eds.}
\begin{document}

\maketitle

\begin{abstract}

Once understood as the paradigm of passively evolving objects, the discovery that massive galaxies
experienced an enormous structural evolution in the last ten billion years has opened an active line of
research. The most significant pending question in this field is the following: which mechanism has made
galaxies to grow largely in size without altering their stellar populations properties dramatically? The most
viable explanation is that massive galaxies have undergone a significant number of minor mergers which have
deposited most of their material in the outer regions of the massive galaxies. This scenario, although
appealing, is still far from be observationally proved since the number of satellite galaxies surrounding the
massive objects appears insufficient at all redshifts. The presence also of a population of nearby massive
compact galaxies with mixture stellar properties is another piece of the puzzle that still does not nicely
fit within a comprehensive scheme. I will review these and other intriguing properties of the massive
galaxies in this contribution.

\end{abstract}

\firstsection % if your document starts with a section,

\section{Introduction}

The discovery that massive galaxies were much more compact in the past (Daddi et al. 2005; Trujillo et al.
2006)  revolutionized our traditional picture of how these objects  have developed with cosmic time. A
monolithic-like scenario, where the bulk of the stellar population as well as the structure of these galaxies
are form in a single dissipative event followed by a passive evolution, is not longer
supported by the observations.

\begin{figure}[b]
\begin{center}
 \includegraphics[width=4in]{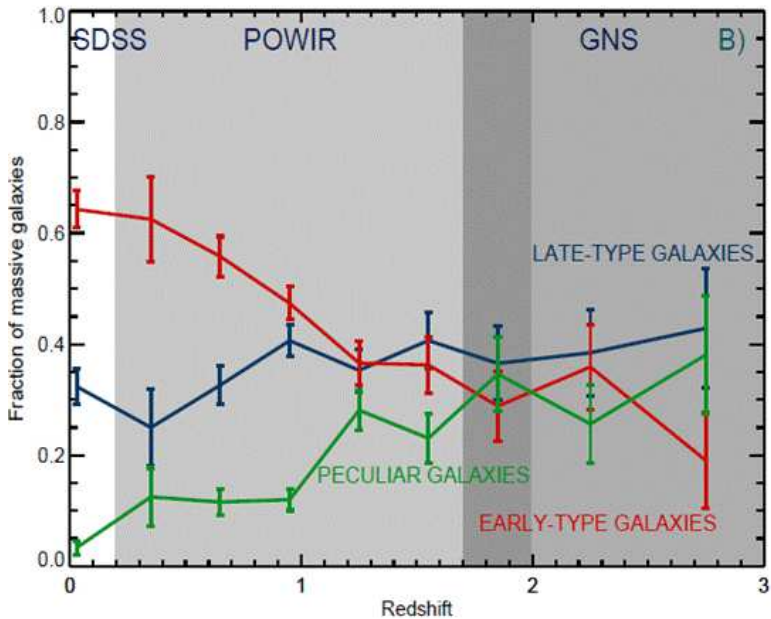} 

 \caption{Fraction of massive (M$_\star$$\sim$10$^{11}$M$_{sun}$) galaxies as function of redshift
segregating the objects according to their visual morphological classification. Blue color represents late
type (S) objects and red early type (E+S0) galaxies, while peculiar (ongoing mergers and irregulars) galaxies
are tagged in green. Different color backgrounds indicate the redshift range
expanded for each survey used: SDSS, POWIR/DEEP2 and GNS. Error bars are estimated following a binomial
distribution. Figure taken from Buitrago et al. (2011).}

   \label{fig0}

\end{center}
\end{figure}

As any shift in scientific paradigm, there has been an enormous debate about the reality of this huge
structural evolution. Most of the critics against this discovery focused on the reliability of the size
estimations  and the accuracy of the stellar mass determinations of these z$>$1 objects (e.g. Mancini et al.
2010; Muzzin et al. 2009). Today, ultra-deep observations of these galaxies  (e.g. Carrasco et al. 2010;
Cassata et al. 2010) as well as the first dynamical estimations of their masses (e.g. Cenarro \& Trujillo 2009;
Cappellari et al. 2009) have inclined the vast majority of the community to accept as real the
size evolution of the massive galaxies. But not only the size of the massive galaxies have dramatically
changed with comic time, also the morphological content among the family of massive galaxies has drastically
varied as redshift decreases (see Fig. \ref{fig0}). In fact, present-day massive galaxies are composed mostly by objects with
spheroidal-like appearance. At high-z, the most common morphology of the massive galaxies resembled disk-like
structures (e.g. van der Wel et al. 2011; Buitrago et al. 2011).

\begin{figure}[b]
\begin{center}
 \includegraphics[width=5in]{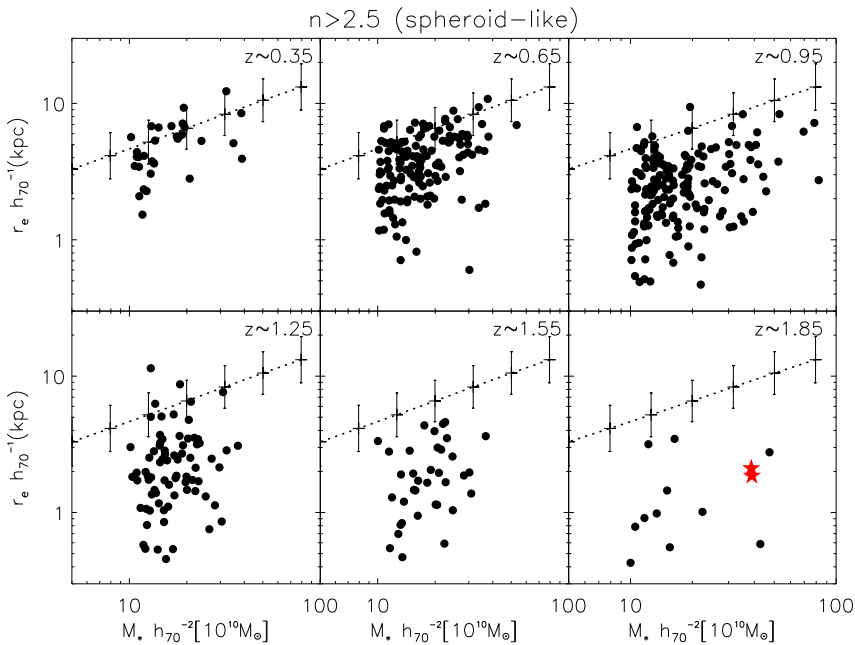} 

 \caption{Stellar mass-size distribution of our high-concentrated (spheroid-like) galaxies. Over-plotted on
the observed distribution of points are the mean and dispersion of the distribution of the S\'ersic
half-light radius of the SDSS early-type (n$>$2.5; Shen et al. 2003) galaxies as a function of the stellar mass. For clarity,
individual error bars are not shown. The mean size relative error is $<$30 per cent. Uncertainties in the
stellar mass are $\sim$0.2 dex. Solid black points are from the ACS sample of Trujillo et al. (2007), red stars
from Carrasco et al. (2010) Gemini high resolution imaging.}

   \label{fig1}

\end{center}
\end{figure}

The stellar mass-size relation of massive galaxies seem to be at place (although with a different "zeropoint"
position than in the present-day universe) since at least z$\sim$3 (e.g. Trujillo et al. 2007; Buitrago et
al. 2008) and the scatter along this relation has not significantly changed since then (see Fig. \ref{fig1}).
However, the number of galaxies that populate these relations have grown with time as the number density of
massive galaxies have continuously increasing since that epoch (e.g. P\'erez-Gonz\'alez et al. 2008). That
means that the new massive galaxies that are incorporated in the stellar mass-size relation are located in
such sense that do not alter dramatically this relation. In order to maintain the scatter of this relation
relatively constant with time, the newcomers should evolve later in size similarly as the older galaxies
that  already populated the stellar mass-size relation.

On what follows I will summarize the different scenarios that have been proposed to explain the significant
structural change of the massive galaxies as well as  the observational evidence favoring the different
mechanisms. We adopt a cosmology with $\Omega_m$=0.3, $\Omega_\Lambda$=0.7 and H$_0$=70 km s$^{-1}$
Mpc$^{-1}$.

\section{What is the physical mechanism behind the size evolution?}

If we accept the reality of the structural evolution of the massive galaxies, the next question to solve is
how these objects have reached their present configuration. We can summarize the different proposed scenarios
in three categories. It is worth stressing that the following mechanisms can take all place simultaneously
and certainly they should all have a role in the evolution of the massive galaxies. Consequently, when we use
the word rejected or supported by the observations we will be referring to the role of such mechanism as the
main driver of the size evolution.

\begin{itemize}

\item Major mergers. This was the earliest theoretical suggestion (e.g. Naab et al. 2007; Nipoti et al. 2010)
and it was also the first hypothesis rejected by the observations (e.g. Bundy et al. 2009; Wild et al. 2009;
de Ravel et al. 2009; Bluck et al. 2009; L\'opez-San Juan et al. 2010). Simply, there is not enough number of
major mergers that can account by the huge size evolution observed (a factor of 4 since z$\sim$2; Trujillo et
al. 2007) plus the relatively modest evolution in stellar mass (a factor of 2 since z$\sim$2; van Dokkum et
al. 2010). The predicted size evolution as a function of the increase in mass goes as:
$\Delta$r$_e$$\propto$$\Delta$M in major mergers (e.g. Ciotti \& van Albada 2001; Boylan-Kolchin et al. 2006)
which is insufficient to produce the observed size evolution.

\item Puffing up. Fan et al. (2008; 2010) as well as Damjanov et al. (2009) proposed a scenario where the size
evolution is connected to the massive expulsion of gas by the effect of an AGN (Fan et al.) or stellar winds
(Damjanov et al.). According to this mechanism, the removal of gas changes the gravitational potential of the galaxy
making the object to puff up to its new (larger) configuration. This evolution is fast ($\lesssim$1 Gyr;
Ragone-Figueroa \& Granato 2011) and the model predicts a dichotomy of massive objects at all redshifts: young ones
($<$1 Gyr) with small sizes and high velocity dispersions ($\sim$400 km/s) and old ones ($>$1 Gyr) with
present-day sizes and moderate velocity dispersion ($\sim$200 km/s). This is not observed in nature: massive
compact galaxies at high-z are "old" at those epochs and there is not an age segregation in the stellar
mass-size relation since, at least, z=1 for objects with spheroid-like morphologies (Trujillo et al. 2011).
Summarizing, this scenario is also not favored observationally.

\item Minor mergers. This model (Khochfar \& Burkert 2006; Maller et al. 2006; Hopkins et al. 2009b; Naab et
al. 2009; Sommer-Larsen \& Toft  2010; Oser et al. 2010) proposes that most of the size evolution of the
massive galaxies has taken place due to the continuous accretion of minor bodies. The stars of these merged
satellites are mainly located in the periphery of the  main body, making this mechanism an excellent vehicle
for the size evolution. The predicted increase in size as a function of the  increase in mass goes as:
$\Delta$r$_e$$\propto$$\Delta$M$^2$ (e.g. Naab et al. 2009). This evolutionary path predicts the following
observables: a continuous increase in size of the global population of massive galaxies, a size growth  not
related with the age of the main galaxy, a mild velocity dispersion evolution of the massive galaxy with
time (Hopkins et al. 2009b).

\end{itemize}

\section{Observational evidence favoring the minor merging scenario}

\begin{figure}[b]
\begin{center}
 \includegraphics[width=5in]{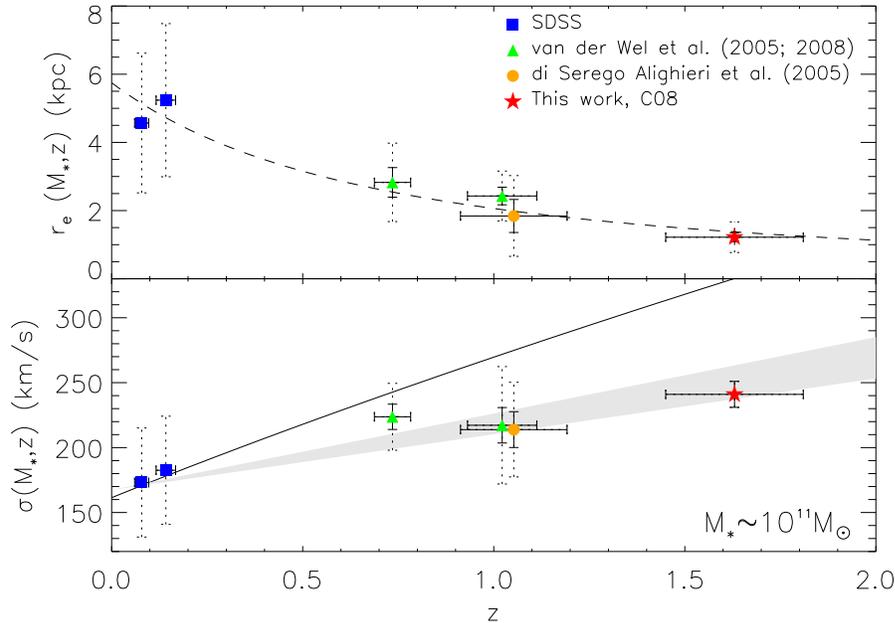} 

 \caption{Top panel: size evolution of M$_\star$$\sim$10$^{11}$M$_{sun}$ spheroid-like galaxies as a function
of redshift. Different symbols show the median values of the effective radii for the different galaxy sets
considered in this work (see Section 3), as indicated in the labels. Dashed error bars, if available, show
the dispersion of the sample, whereas the solid error bars indicate the uncertainty of the median value. The
dashed line represents the observed evolution of sizes r$_e$(z)$\propto$(1 + z)$^{-1.48}$ found in Buitrago
et al. (2008; B08) for galaxies of similar stellar mass. Bottom panel: velocity dispersion evolution of the
spheroid-like galaxies as a function of redshift, with symbols as given above. Assuming the B08 size
evolution, the solid line represents the prediction from the "puffing-up" scenario (Fan et al.
2008), whereas the gray area illustrates the velocity dispersion evolution within the merger scenario of
Hopkins et al. (2009b) for 1$<$$\gamma$$<$2. Figure from Cenarro \& Trujillo (2009).}

   \label{fig2}

\end{center}
\end{figure}

There are many observational evidences favoring the minor merging hypothesis as the main channel of massive
galaxies growth. We can summarize them in three groups:

\begin{itemize}

\item The size evolution of the spheroid-like massive galaxies is not related with the age of their stellar
population. Since z$\sim$1, spheroid-like massive galaxies, at a given fixed stellar mass,  still need to
grow by a factor $\sim$2 to reach their present configuration. This significant size evolution is observed, at
all redshifts, to be independent of the stellar age of the massive galaxies (Trujillo et al. 2011). This
observation points out to a size growth mechanism that does not know about the age of the main galaxy. An
external accretion of stars (where the infalling satellites do not have previous knowledge about the age of
the central galaxy) fits well within this scheme.

\item There is a progressive and steady formation of the outer galaxy envelopes. The central stellar mass
density of the massive galaxies at high-z do not dramatically differ from the central stellar mass density of
the nearby massive galaxies (Bezanson et al. 2009; Hopkins et al. 2009a). The majority of the evolution of the
stellar mass density profile of the massive galaxies has taken place at their extended wings. Massive
galaxies have steadily increased their number of stars at farther distances (van Dokkum et al. 2010). This
progressive build is very suggestive of a continuous accretion of new stars with cosmic time in the
periphery of these galaxies.

\item At a fixed stellar mass, the velocity dispersion of the massive galaxies has mildly declined since
z$\sim$2. Cenarro \& Trujillo (2009) compiled from the literature the velocity dispersions of many massive
(M$_\star$$\sim$10$^{11}$M$_{sun}$) galaxies since z$\sim$2. This compilation took data from van der Wel et
al (2005; 2008) at 0.5$<$z$<$1 and di Seregho Allighieri et al. (2005) at z$\sim$1. This data was
complemented with the measurement of the velocity dispersions of massive galaxies  in the SDSS (for having a
local reference) and with the first estimation of the velocity dispersion of massive galaxies at z$>$1.5
(using the published staked spectrum of Cimatti et al. 2008). All this data together (see Fig. \ref{fig2})
clearly indicated that the evolution of the velocity dispersion of the massive galaxies, at a fixed stellar
mass, has only moderately declined with cosmic time. This result has been later confirmed by many new
estimations of the velocity dispersion of massive galaxies at 1.5$<$z$<$2 (e.g. Cappellari et al. 2009;
Onodera et al. 2010; van de Sande et al 2011; Newman et al. 2010, Toft et al. 2012). This mild evolution of
the velocity dispersion is in good agreement with the idea that most of the structural evolution of the
massive galaxies has taken place in their outer regions. This again fits well with a scenario of accretion of
new stars that is smooth and mostly locate stars in the periphery of these objects. The fact that the central
stellar mass density of the massive galaxies has only changed mildly since z$\sim$2 also agrees with the fact
that the central velocity dispersion of these objects have not changed significantly (see a much elaborated
discussion of this point in Trujillo et al. 2012). 

\end{itemize} 

\section{Some puzzling observations}

So far, both the cosmological simulations as well as the observational evidence favor the minor merging
scenario as the main driver of size and mass evolution of the massive galaxies. However, there are two
observational evidences that are not easy to understand, at least with the present theoretical development,
within the minor merging hypothesis. These two puzzling observations are: the scarcity of massive
compact galaxies in the local universe and the factor of 2 less satellites surrounding the massive galaxies
at all redshifts compared with the model predictions.

\subsection{Nearby massive compact galaxies: relics of the early universe?}

\begin{figure}[b]
\begin{center}
 \includegraphics[width=5in]{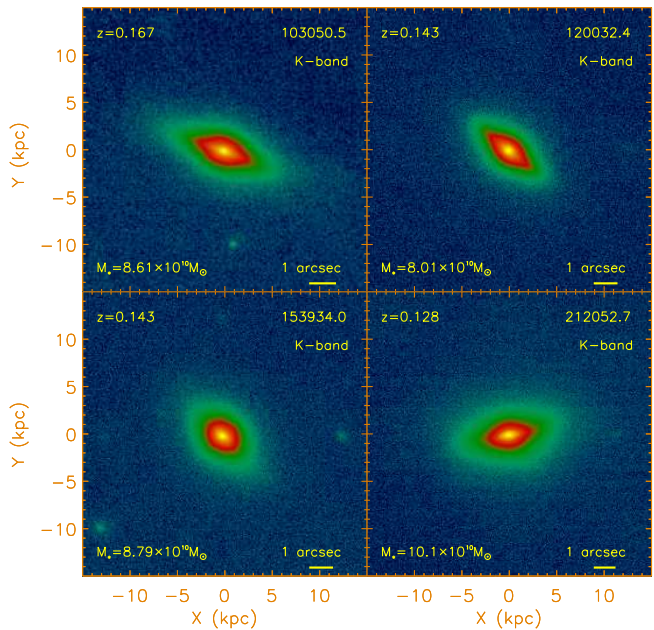} 

 \caption{K-band Gemini high-resolution (FWHM$\sim$0.2 arcsec) imaging of four nearby (z$\sim$0.15)
massive compact galaxies. Listed on each figure is the galaxy name, its stellar mass, and its spectroscopic
redshift. The solid line indicates 1 arcsec angular size. Figure taken from Trujillo et al. (2012).}

   \label{fig3}

\end{center}
\end{figure}

After the discovery that massive galaxies at high-z were compact, there was an observational effort to try
finding massive  (M$_\star$$\sim$10$^{11}$M$_{sun}$) and compact (r$_e$$\sim$1 kpc) objects in the nearby
Universe (see  Fig. \ref{fig3}). According to the theoretical predictions (Hopkins et al. 2009b) around 10\%
of the massive compact galaxies since z$\sim$2 should have survived intact  due to the stochastic nature of
the merging channel. Taking into account that the number density of massive galaxies at z$\sim$2  was a
factor of 10 smaller than today, around 1\% of the present-day massive
galaxy population should be composed by relics (i.e. they should appear today as old compact massive galaxies) from that early epoch
of the Universe. Observationally, it is found that less than 0.03\% of the current massive galaxies are as
compact as the ones found at z$\sim$2 (Trujillo et al. 2009). Moreover, these galaxies are not only very
scarce (see also Taylor et al. 2010) but young ($\sim$2 Gyr; Ferr\'e-Mateu et al. 2012). Consequently, it
seems that massive compact relics of the early universe are non-existent today. 

Valentinuzzi et al. (2010) have argued against the above findings and claim that old and dense massive
galaxies can be found in large numbers in galaxy clusters environments. This will alleviate the problem for
minor merging scenarios. However, it is difficult to understand why these nearby massive compact objects have
not popped out in the SDSS survey (which contain many of these galaxy clusters). So, the controversy
still remains open and further investigation is required.

\subsection{Satellites surrounding massive galaxies: are they enough?}

If the minor merging scenario is the main channel of massive galaxy evolution one would expect that a direct
test of this hypothesis could be done by counting the number of satellites surrounding these galaxies and
exploring their evolution with redshift. Estimating the number of satellites around massive galaxies have
been done by many authors (Kaviraj et al. 2009; Jackson et al. 2010; Nierenberg et al. 2011;  Man et al.
2012; Newman et al. 2012; M\'armol-Queralt\'o et al.2012a). As expected, the fraction of massive galaxies with
nearby satellites depends on two parameters: the search radius to find the satellite and the mass ratio
between the massive galaxy and the satellite. To give a number, Liu et al. (2011) found that around 13\% of
the local galaxies with M$_\star$$\gtrsim$10$^{11}$M$_{sun}$ have a satellite with a mass ratio
1:10 or smaller within a projected radius of 100 kpc to the host galaxy. This fraction is constant with
redshift (M\'armol-Queralt\'o et al. 2012a; at least up to z$\sim$2). If the explored mass ratio is decreased
down to 1:100, the fraction of massive galaxies with a nearby satellite increases up to $\sim$30\%.

\begin{figure}[b]
\begin{center}

\vspace{0.5cm}

 \includegraphics[width=5in]{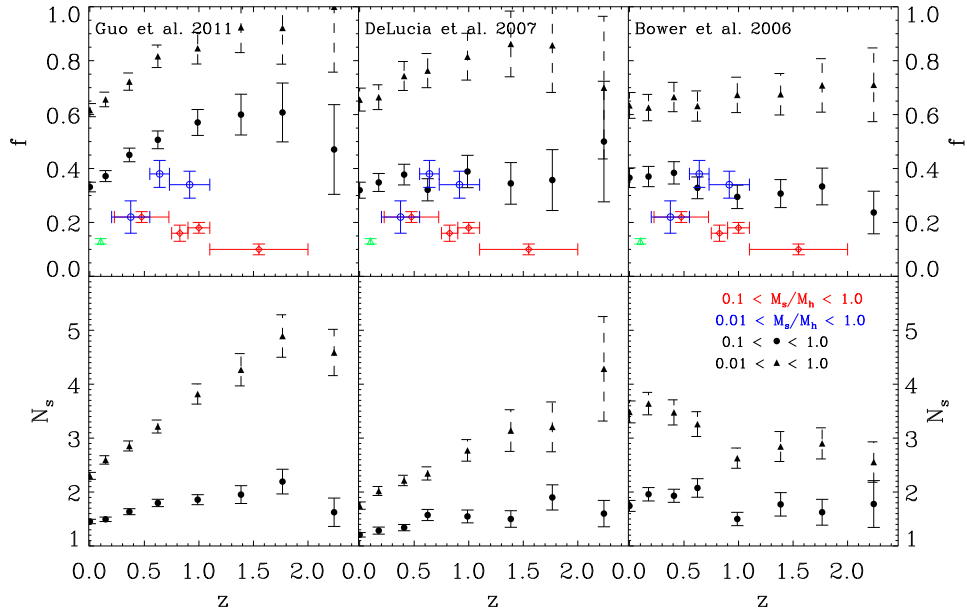} 

 \caption{Columns stand for the results of three galaxy catalogs based on different semi-analytical models.
For each model, and from top to bottom, are shown the fraction of massive galaxies that have at least one
satellite within a sphere of 100 kpc radius and a projected distance smaller than 100 kpc, and the average
number of satellites per massive galaxy when they have one of such objects around. The full circles
(triangle) stand for the satellites with stellar mass ratios of 0.1$<$M$_s$/M$_h$$<$1
(0.01$<$M$_s$/M$_h$$<$1). The error bars represent one standard deviation. The observational data from
M\'armol-Queralt\'o et al. (2012) are overplotted as red (blue) open circles (diamonds) for mass ratios of
0.1$<$M$_s$/M$_h$$<$1.0 (0.01$<$M$_s$/M$_h$$<$1). The local observational reference (z=0.1) from Liu et al.
(2011) for the fraction of massive galaxies with satellites with mass ratios of 0.1$<$M$_s$/M$_h$$<$1 is
plotted as a green open triangle, no data are available for smaller satellites. Figure taken from Quilis \&
Trujillo (2012).}

   \label{fig4}

\end{center}
\end{figure}

The above fractions can be directly compared with the predictions from  $\Lambda$CDM cosmological
simulations. In particular, these numbers can be confronted with the semianalytical predictions based on
those simulations. This simply exercise was conducted by Quilis \& Trujillo (2012) using three different
semianalytical models (Bower et al. 2006; De Lucia \& Blaizot 2007 and Guo et al. 2011) run over the
Millenium I (Springel et al. 2005) and Millenium II (Boylan-Kolchin et al. 2009) simulations. Interestingly, the theoretical models predicted correctly the constancy on the
fraction of massive galaxies with nearby satellites across the cosmic time. However, all models
overpredicted by a factor of $\sim$2 the value of this fraction (see Fig. \ref{fig4}). In other words, in the simulations there is
an excess of satellites that could  later  merge with the massive host galaxy. Whether this excess of
satellites in the simulations is also overpredicting the size evolution that is obtained in the simulations
is a matter of analysis at the moment of writing this review.

\section{Open questions}

There are a number of observational predictions associated to the minor merging scenario that could be tested
with new observations. One of these predictions is a significant radial change in the stellar population
properties of the massive galaxies. In particular, one would expect an age and metallicity gradient if the
infalling satellites have progressively form the outer parts of the massive galaxies. In this scheme, the
outer regions of the massive galaxies should be progressively young and metal poor.

Age gradients should be extremely difficult to measure observationally in present-day massive galaxies. Our
ability to distinguish among a few Gyr different old stellar populations is very limited at present. The
stars accreted through the infalling satellites should be relatively evolved at the moment of the merger and,
after the infall to the main galaxy, no new star formation is expected to occur. Consequently, we should not
expect many evidence of  minor merging happened at z$\gtrsim$0.5 exploring age gradients of nearby massive
galaxies. However, metallicity gradients should be more easily identified in present-day massive galaxies as
the metallicities of the stellar populations should not evolve with cosmic time. 

There are a few but increasing number of studies exploring age and metallicity radial gradients in nearby
massive galaxies up to several effective radii (Coccato et al. 2010; Tal et al. 2011; Greene et al. 2012; La
Barbera et al. 2012). These studies agree on a metallicity decrease of the stellar populations of massive
galaxies towards the outer regions. These works, however, are still at their infancy as measuring the stellar
population properties at such distances is complicated due to the low surface brightness of the stellar
populations at those radii. 

Another further advance that is expected in the next few years is the measurement of the stellar population
properties of the satellite galaxies that will eventually merge with the massive galaxies. This information
is key if we want to close the loop with the evidence compiled from the outer parts of present-day massive
galaxies. A strong consistency test for the minor merging scenario is that the information provided by both
types of works agrees. A few studies have pioneered the analysis of the stellar populations of satellite
galaxies at high-z (Newman et al. 2012; M\'armol-Queralt\'o et al. 2012b). High-z satellites had similar ages
than their massive hosts but that changed with time, and present-day satellites are much younger than their
massive galaxy. Measuring metallicities of the satellite galaxies at high-z has not been conducted yet. A
step further in this sense it is expected with the new SHARDS survey (P\'erez-Gonz\'alez et al. 2012) 
 conducted with the 10.4 m GTC telescope at La Palma. 

\section{Summary}

The discovery that massive galaxies were much more compact in the past has opened a fruitful era of research
 trying to put this finding within a galaxy formation context. Both theory and
observations seem to converge to a scenario where the main channel of size and mass evolution of the massive
galaxies is through a continuous accretion of minor bodies as cosmic time progresses. This active life of the
massive galaxies  follows after a rapid (dissipative) collapse which would have form the bulk of the present-day
body of the most massive objects (see e.g. Dekel et al. 2009; Ricciardelli et al. 2010; Wuyts et al. 2010;
Bournaud et al. 2011;  Targett et al. 2011; Barro et al. 2012). 

Although the general picture of massive galaxy evolution seems to be at place, still a few observational
results challenge this scenario: the nearly absence of old compact massive relics in the present-day universe
and the apparent few satellites that surround the massive galaxies at every redshift. Further investigations
will clarify whether this discrepancy is just a matter of refining the models predictions or whether these
observations will force us to change our main view of massive galaxy evolution.

\acknowledgments

The results presented here are due to the effort of many people over the last 5 years. I would like to
particularly thank the large number of collaborators which I have had the pleasure to work with along all
these years. This work has been supported by the Programa Nacional de Astronom\'ia y Astrof\'isica of the
Spanish Ministry of Science and Innovation under grant AYA2010-21322-C03-02.


\begin{thebibliography}{}

\bibitem[barro2012]{barro2012}
{Barro, G. et al.} 2012,
\textit{ApJ} (Letters), submitted, arXiv:1206.5000

\bibitem[bezanson2009]{bezanson2009}
{Bezanson R., van Dokkum P. G., Tal T., Marchesini D., Kriex M., \& Coppi P.} 2009,
\textit{ApJ}, 697, 1290

\bibitem[bluck2009]{bluck2009}
{Bluck, A. F. L., Conselice, C. J., Bouwens, R. J., Daddi, E., Dickinson, M., Papovich, C., \& Yan, H.} 2009,
\textit{MNRAS} (Letters), 394, L51

\bibitem[bournaud2011]{bournaud2011}
{Bournaud, F. et al.} 2011,
\textit{ApJ}, 730, 4

\bibitem[bower2006]{bower2006}
{Bower, R. G., Benson, A. J., \& Malbon, R. et al.} 2006, 
\textit{MNRAS}, 370, 645

\bibitem[boylan2006]{boylan2006}
{Boylan-Kolchin M., Ma C.-P., \& Quataert E.} 2006, 
\textit{MNRAS}, 369, 1081

\bibitem[boylan2009]{boylan2009}
{Boylan-Kolchin, M., Springel, V., White, S. D. M., Jenkins, A., \& Lemson, G.} 2009,
\textit{MNRAS}, 398, 1150

\bibitem[buitrago2008]{buitrago2008}
{Buitrago, F., Trujillo, I., Conselice, C. J., Bouwens, R. J., Dickinson, M., \& Yan, H.} 2008,
\textit{ApJ} (Letters), 687, L61

\bibitem[buitrago2012]{buitrago2012}
{Buitrago, F., Trujillo, I., Conselice, C. J., Haeussler, B.}
\textit{MNRAS}, in press, arXiv:1111.6993

\bibitem[bundy2009]{bundy2009}
{Bundy K., Fukugita M., Ellis R. S., Targett T. A., Belli S., \& Kodama T.} 2009, 
\textit{ApJ}, 697, 1369

\bibitem[cappellari2009]{cappellari2009}
{Cappellari, M et al.} 2009,
\textit{ApJ} (Letters), 704, L34

\bibitem[carrasco2010]{carrasco2010}
{Carrasco, E. R., Conselice, C. J., \& Trujillo, I.} 2010,
\textit{MNRAS}, 405, 2253

\bibitem[cassata2010]{cassata2010}
{Cassata, P. et al.} 2010,
\textit{ApJ} (Letters), 714, L79

\bibitem[centru2009]{centru2009}
{Cenarro, A. J., \& Trujillo, I.} 2009,
\textit{ApJ} (Letters), 696, 43

\bibitem[cimatti2008]{cimatti2008}
{Cimatti A. et al.} 2008, 
\textit{A\&A}, 482, 21

\bibitem[ciotti2001]{ciotti2001}
{Ciotti L., \& van Albada T. S.} 2001, 
\textit{ApJ} (Letters), 552, L13

\bibitem[coccato2010]{coccato2010}
{Coccato, L., Gerhard, O., \& Arnaboldi, M.} 2010,
\textit{MNRAS} (Letters), 407, 26

\bibitem[daddi2005]{daddi2005}
{Daddi, E. et al.} 2005,
\textit{ApJ}, 626, 680

\bibitem[damjanov2009]{damjanov2009}
{Damjanov I. et al.} 2009, 
\textit{ApJ}, 695, 101

\bibitem[dekel2009]{dekel2009}
{Dekel, A., et al.} 2009,
\textit{Nature}, 457, 451

\bibitem[delucia2007]{delucia2007}
{De Lucia, G., \& Blaizot, J.} 2007, 
\textit{MNRAS}, 375, 2

\bibitem[deravel2009]{deravel2009}
{de Ravel L., Le F\`evre O., Tresse L. et al.} 2009, 
\textit{A\&A}, 498, 379

\bibitem[diseregho2005]{diseregho2005}
{di Serego Alighieri, S. et al.} 2005,
\textit{A\&A}, 442, 125

\bibitem[fan2008]{fan2008}
{Fan L., Lapi A., De Zotti G., \& Danese L.} 2008, 
\textit{ApJ} (Letters), 689, L101

\bibitem[fan2010]{fan2010}
{Fan L., Lapi A., Bressan A., Bernardi M., De Zotti G., \& Danese L.} 2010, 
\textit{ApJ}, 718, 1460

\bibitem[ferremateu2012]{2012}
{Ferr\'e-Mateu, A., Vazdekis, A., Trujillo, I., S\'anchez-Bl\'azquez, P., Ricciardelli, E., \& de la Rosa, I.
G.} 2012,
\textit{MNRAS}, 423, 632 

\bibitem[greene2012]{greene2012}
{Greene, J. E., Murphy, J. D., Comerford, J. M., Gebhardt, K., \& Adams, J. J.} 2012,
\textit{ApJ}, 750, 32

\bibitem[guo2011]{guo2011}
{Guo, Q., White, S., \& Boylan-Kolchin, M. et al.} 2011, 
\textit{MNRAS}, 413, 101 

\bibitem[hopkins2009a]{hopkins2009a}
{Hopkins P. F., Bundy K., Murray N., Quataert E., Lauer T. R., \& Ma. C.} 2009a,
\textit{MNRAS}, 398, 898

\bibitem[hopkins2009b]{hopkins2009b}
{Hopkins P. F., Hernquist L., Cox T. J., Keres D., \& Wuyts S.} 2009b, 
\textit{ApJ}, 691, 1424

\bibitem[khochfar2006]{khochfar2006}
{Khochfar S., \& Burkert A.} 2006,
\textit{A\&A}, 445, 403

\bibitem[labarbera2012]{labarbera2012}
{La Barbera, F., Ferreras, I., de Carvalho, R. R., Bruzual, G., Charlot, S., Pasquali, A., \& Merlin, E.}
2012,
\textit{MNRAS}, in press, arXiv:1208.0587

\bibitem[liu2011]{liu2011}
{Liu, L., Gerke, B. F., Wechsler, R. H., Behroozi, P. S., \& Busha, M. T.} 2011,
\textit{ApJ} (Letters), 733, L62

\bibitem[lopezsanjuan2010]{lopezsanjuan2010}
{L\'opez-Sanjuan C., Balcells M., P\'erez-Gonz\'alez P. G., Barro G., Garc\'ia-Dab\'o C. E., Gallego J., \&
Zamorano J.} 2010, 
\textit{ApJ}, 710, 1170

\bibitem[maller2006]{maller2006}
{Maller A. H., Katz N., Keres D., Dav\'e R., \& Weinberg D. H.} 2006, 
\textit{ApJ}, 647, 763

\bibitem[man2012]{man2012}
{Man, A. W. S., Toft, S., Zirm, A. W., Wuyts, S., \& van der Wel, A.} 2012,
\textit{ApJ}, 744, 85

\bibitem[mancini2010]{mancini2010}
{Mancini, C. et al.} 2010
\textit{MNRAS}, 401, 933

\bibitem[marmol2012a]{marmol2012a}
{M\'armol-Queralt\'o, E., Trujillo, I., P\'erez-Gonz\'alez, P. G., Varela, J., \& Barro, G.} 2012a,
\textit{MNRAS}, 422, 2187

\bibitem[marmol2012b]{marmol2012b}
{M\'armol-Queralt\'o, E., et al.} 2012b,
\textit{MNRAS}, submitted

\bibitem[muzzin2009]{muzzin2009}
{Muzzin, A., van Dokkum, P., Franx, M., Marchesini, D., Kriek, M., Labb\'e, I.} 2009,
\textit{ApJ} (Letters), 706, L188

\bibitem[naab07]{naab07}
{Naab T., Johansson P. H., Ostriker J. P., \& Efstathiou G.} 2007, 
\textit{ApJ}, 658, 710

\bibitem[naab09]{naab09}
{Naab T., Johansson P. H., \& Ostriker J. P.} 2009, 
\textit{ApJ} (Letters), 699, L178

\bibitem[newman2010]{newman2010}
{Newman, A. B., Ellis, R. S., Treu, T., Bundy, K.} 2010,
\textit{ApJ} (Letters), 717, L103

\bibitem[newman2012]{newman2012} 
{Newman, A. B., Ellis, R. S., Bundy, K., Treu, T.} 2012,
\textit{ApJ}, 746, 162

\bibitem[nierenberg2011]{nierenberg2011}
{Nierenberg, A. M., Auger, M. W., Treu, T., Marshall, P. J., Fassnacht, C. D.} 2011,
\textit{ApJ}, 731, 44

\bibitem[nipoti2010]{nipoti2010}
{Nipoti C., Londrillo P., \& Ciotti L.} 2003, 
\textit{MNRAS}, 342, 501

\bibitem[onodera2010]{onodera2010}
{Onodera, M. et al.} 2010,
\textit{ApJ} (Letters), 715, L6O

\bibitem[oser2010]{oser2010}
{Oser, L., Ostriker, J. P., Naab, T., Johansson, P. H., Burkert, A.} 2010,
\textit{ApJ}, 725, 2312

\bibitem[perezgonzalez2008]{perezgonzalez2008}
{P\'erez-Gonz\'alez, P. G. et al.} 2008,
\textit{ApJ}, 675, 234

\bibitem[perezgonzalez2012]{perezgonzalez2012}
{P\'erez-Gonz\'alez, P. G. et al.} 2012,
\textit{ApJ}, in press, arXiv:1207.6639

\bibitem[quilis2012]{quilis2012}
{Quilis, V., \& Trujillo, I.} 2012,
\textit{ApJ} (Letters), 752, L19

\bibitem[ragonefigueroa2011]{ragonefigueroa2011}
{Ragone-Figueroa C., \& Granato G. L.} 2011, 
\textit{MNRAS}, 414, 3690

\bibitem[ricciardelli2010]{ricciardelli2010}
{Ricciardelli, E., Trujillo, I., Buitrago, F., \& Conselice, C. J.} 2010,
\textit{MNRAS}, 406, 230

\bibitem[shen2003]{shen2003}
{Shen S., Mo H. J., White S. D. M., Blanton M. R., Kauffmann G., Voges W., Brinkmann J., Csabai I.} 2003,
\textit{MNRAS}, 343, 978

\bibitem[sommerlarsen2010]{sommerlarsen2010}
{Sommer-Larsen, J., \& Toft, S.} 2010,
\textit{ApJ}, 721, 1755

\bibitem[springel2005]{springel2005}
{Springel, V., White, S. D. M., \& Jenkins, A. et al.} 2005, 
\textit{Nature}, 435, 629 

\bibitem[tal2011]{tal2011}
{Tal, T., van Dokkum, P. G.} 2011,
\textit{ApJ}, 731, 89

\bibitem[targett2011]{targett2011}
{Targett, T. A., Dunlop, J. S., McLure, R. J., Best, P. N., Cirasuolo, M., \& Almaini, O.} 2011,
\textit{MNRAS}, 412, 295

\bibitem[taylor2010]{taylor2010}
{Taylor E. N., Franx M., Glazebrook K., Brinchmann J., van der Wel A., \& van Dokkum P. G.} 2010,
\textit{ApJ}, 720, 723

\bibitem[toft2012]{toft2012}
{Toft, S., Gallazzi, A., Zirm, A., Wold, M., Zibetti, S., Grillo, C., Man, A.} 2012,
\textit{ApJ}, 754, 3

\bibitem[truji2006]{truji2006}
{Trujillo, I. et al.} 2006,
\textit{MNRAS} (Letters), 373, L36

\bibitem[truji2007]{truji2007}
{Trujillo, I., Conselice, C. J., Bundy, K., Cooper, M. C., Eisenhardt, P., \& Ellis, R. S.} 2007,
\textit{MNRAS}, 382, 109 

\bibitem[truji2009]{truji2009}
{Trujillo I., Cenarro A. J., de Lorenzo-C\'aceres A., Vazdekis A., de la Rosa I. G., \& Cava A.} 2009, 
\textit{ApJ} (Letters), 692, L118

\bibitem[truji2011]{truji2011}
{Trujillo, I., Ferreras, I., \& de La Rosa, I. G.} 2011,
\textit{MNRAS}, 415, 3903 

\bibitem[truji2012]{truji2012}
{Trujillo, I., Carrasco, E. R., Ferr\'e-Mateu, A.} 2012,
\textit{ApJ}, 751, 45

\bibitem[valentinuzzi2010]{valentinuzzi2010}
{Valentinuzzi T. et al.} 2010, 
\textit{ApJ}, 712, 226

\bibitem[vanderwel2005]{vanderwel2005}
{van der Wel, A., Franx, M., van Dokkum, P. G., Rix, H.-W., Illingworth, G. D., \& Rosati, P.} 2005,
\textit{ApJ}, 631, 145

\bibitem[vanderwel2008]{vanderwel2008}
{van der Wel A., Holden B. P., Zirm A. W., Franx M., Rettura A., Illingworth G. D., \& Ford H. C.} 2008,
\textit{ApJ}, 688, 48

\bibitem[vanderwel2011]{vanderwel2011}
{van der Wel A., et al.} 2011,
\textit{ApJ}, 730, 38

\bibitem[vandesande2011]{vandesande2011}
{van de Sande, J. et al.} 2011,
\textit{ApJ} (Letters), 736, L9

\bibitem[vandokkum2010]{vandokkum2010}
{van Dokkum P. G. et al.} 2010, 
\textit{ApJ}, 709, 1018

\bibitem[wild2009]{wild2009}
{Wild V., Walcher C. J., Johanson P. H., Tresse L., Charlot S., Pollo A., Le F\`evre O., \& de Ravel L.} 2009, 
\textit{MNRAS}, 395, 144

\bibitem[wuyts2010]{wuyts2010}
{Wuyts, S., Cox, T. J., Hayward, C. C., Franx, M., Hernquist, L., Hopkins, P. F., Jonsson, P., van Dokkum, P.
G.} 2010,
\textit{ApJ}, 722, 1666

\end{thebibliography}
\end{document}